\documentclass[twocolumn,preprintnumbers,amsmath,amssymb,prl,footinbib]{revtex4}

\usepackage{graphicx}
\usepackage{epsfig}
\usepackage{epsf}
\usepackage{amssymb}
\usepackage{mathrsfs}
\usepackage{dcolumn}
\usepackage{amsbsy}
\usepackage{bm}
\usepackage{color} 
\usepackage{subfigure}
\usepackage{setspace}
\usepackage{multirow}
\usepackage{ulem} 

\begin{document}
\title{Universal Control of Nuclear Spins Via Anisotropic Hyperfine Interactions}
\author{J. S. Hodges, J. C. Yang, C. Ramanathan, D. G. Cory}
\affiliation{Department of Nuclear Science and
Engineering, Massachusetts Institute of Technology, Cambridge, MA 02139, USA}
\date{\today}

\newcommand{\half}{\frac{1}{2}}
\newcommand{\ket}[1]{\vert{#1}\rangle}
\newcommand{\bra}[1]{\langle{#1}\vert}
\newcommand{\ham}{\mathscr{H}}
\newcommand{\Tr}[1]{Tr\{{#1}\}}
\newcommand{\pot}{\frac{\pi}{2}}


\begin{abstract}We show that nuclear spin subsystems can be completely controlled via microwave irradiation of resolved anisotropic hyperfine interactions with a nearby electron spin.  Such indirect addressing of the nuclear spins via coupling to an electron allows us to create nuclear spin gates whose operational time is significantly faster than conventional direct addressing methods.  We experimentally demonstrate the feasibility of this method on a solid-state ensemble system consisting of one electron and one nuclear spin.
\end{abstract}
\maketitle


Coherent control of quantum systems promises optimal computation \cite{Shor}, secure communication \cite{BB84}, and new insight into the fundamental physics of many-body problems \cite{FeynmanQC}.  
Solid-state proposals \cite{KaneQIP,LossDivincenzo,YamamotoSiliconQIP,forschritte,SuterScalableQIP02} for such quantum information processors employ isolated spin degrees of freedom  which provide Hilbert spaces with long coherence times. Here we show how to exploit a local, isolated electron spin to coherently control nuclear spins. Moreover, we suggest that this approach provides a fast and reliable means of controlling nuclear spins and enables the electron spins of such solid-state systems to be used for  state preparation and readout \cite{MehringSpinBus} of nuclear spin states, and \textit{additionally} as a spin actuator for mediating nuclear-nuclear spin gates.

\textbf{Model System.}
The spin Hamiltonian of a single local electron spin with angular momentum, $S = \half$ and N nuclear spins, each with angular momentum $I_k = \half$, in the presence of a magnetic field $\vec{B}$ is \cite{SchweigerBook}:

\begin{eqnarray}
\ham_0 & = & \beta_e \mathbf{g}_{\mu \nu} \hat{S}_\mu B_\nu - \sum_{k=1}^N \gamma_n^k (1-\boldsymbol{\delta}_{\mu\nu}^k) \hat{I}^k_\mu B_\nu \nonumber \\
&&+ 2 \pi \sum_{k=1}^N \mathbf{A}_{\mu \nu}^k \hat{S}_\mu \hat{I}_\nu^k + \pi \sum_{k,l=1}^N \mathbf{D}_{\mu \nu}^{kl} \hat{I}_\mu^k \hat{I}_\nu^l 
\end{eqnarray}
Here $\beta_{e}$ is the Bohr  magneton, $\gamma_n^k$ is the gyromagnetic ratio; $\hat{S}$ and $\hat{I^k}$ are the spin-$\half$ operators.  The second-rank tensors $\mathbf{g}$, $\mathbf{A}^k$, $\boldsymbol{\delta}$, and $\mathbf{D}^{kl}$ represent the electron g-factor, the hyperfine interaction, the chemical shift, and the nuclear dipole-dipole interaction  respectively. 

%
\begin{figure}[htbp] 
   \centering
   \includegraphics[width=2in]{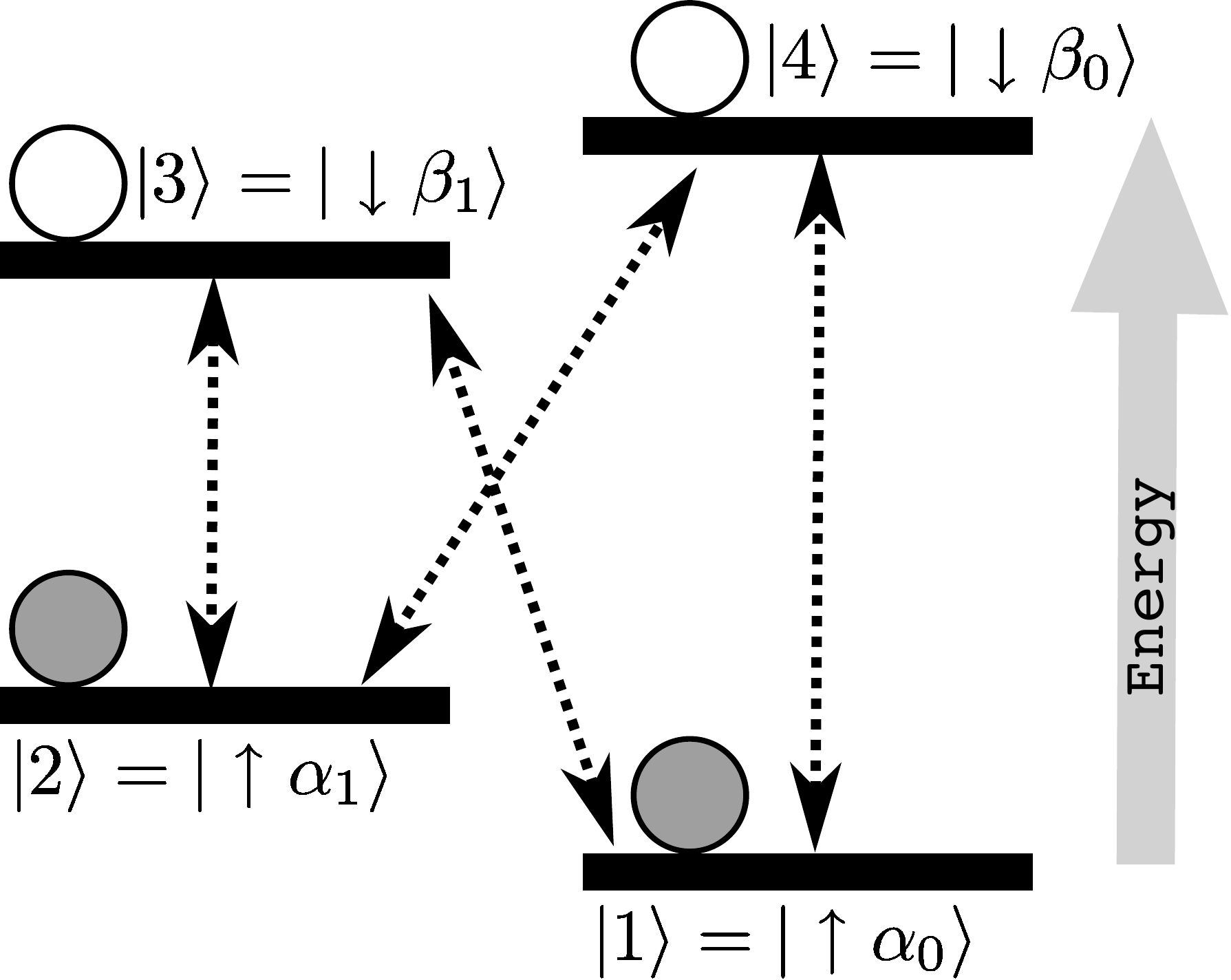} 
   \caption{Energy level diagram for Eq. \eqref{HSimple} with N=1.  The electron spin state is in an eigenstate of purely the Zeeman interaction, while the nuclear spin state is \textit{not} an eigenfunction of the Zeeman interaction alone due to the anisotropic hyperfine interaction.  Because $\langle \alpha_0 \vert \beta_1 \rangle \neq 0$ and  $\langle \alpha_0 \vert \beta_0 \rangle \neq 0$ the electron spin operator ($\hat{S}_x$) has finite  probabilities between all levels (dashed arrows).  This allows for universal control of the entire spin system.  The filled and unfilled circles represent the relative spin state populations of the  ensemble at thermal equilibrium.     In our experimental setup the energy differences are $\omega_{12}/2 \pi = 7.8$ MHz, $\omega_{34}/2 \pi = 40$ MHz, $\omega_{14}/2 \pi = 12.005$ GHz, $\omega_{23}/2 \pi = 11.954$GHz 
  }
   \label{fig:energy}
\end{figure}

In the regime where the static magnetic field $\vec{B} = B_0 \hat{z}$ provides a good quantization axis for the electron spin, the Hamiltonian can be simplified by dropping the non-secular terms which corresponds to keeping only electron interactions involving $S_z$.  The quantization axis of any nuclear spin depends on the magnitudes of the hyperfine interaction and the main magnetic field, as well as their relative orientations.  When these two fields are comparable in magnitude \footnote{The relative order of terms for equation \eqref{HSimple} to hold is: $\beta_e \lVert g_{\mu z} \rVert B_0 \gg \lVert \gamma_n (1-\boldsymbol{\delta}) B_0 \lVert \sim \lVert A_{z\nu} \rVert  \gg \lVert D^{j,k} \rVert$} $\ham_0$ can be approximated by:

\begin{eqnarray}
\label{HSimple}
\ham_0 & \approx & \beta_e  g_{zz} B_0 \hat{S}_z - \sum_{k=1}^N \gamma_n^k (1-\delta_{zz}^k) B_0 \hat{I}^k_z  \nonumber \\ 
& &+ 2\pi \sum_{k=1}^N \Big({A}_{zx}^k \hat{S}_z \hat{I}_x^k +
{A}_{zy}^k \hat{S}_z \hat{I}_y^k +
{A}_{zz}^k \hat{S}_z \hat{I}_z^k 
\Big)
\end{eqnarray}
The nuclear dipole-dipole interaction is neglected as it is typically $\text{10}^\text{2}$ times weaker than the hyperfine terms.

As described in Figure \ref{fig:energy} (N=1), the nuclear spin is quantized in an effective field that is the vector sum of the Zeeman and hyperfine interaction.  The nuclear spin eigenstates are a mixture of the nuclear Zeeman eigenstates: $\ket{\alpha_0} = \cos{\theta_\uparrow}\ket{\uparrow} + e^{i\varphi_k} \sin{\theta_\uparrow} \ket{\downarrow}$.  $\ket{\alpha_1}$ is orthogonal to $\ket{\alpha_0}$.

An N-bit string of quantum information can be stored in this system in either the spin-up or spin-down electron manifolds:
\begin{eqnarray}
\ket{ j_1\; j_2 \; \dots j_N} & = &\ket{\uparrow}\otimes \ket{\alpha_{j_1}^1} \otimes  \ket{\alpha_{j_2}^2} \otimes \dots \otimes  \ket{\alpha_{j_N}^N} \\
& = & \ket{\downarrow}\otimes \ket{\beta_{j_1}^1} \otimes  \ket{\beta_{j_2}^2} \otimes \dots \otimes  \ket{\beta_{j_N}^N}
\end{eqnarray}
where $j_k \in \{0,1\}$.  Note that by storing information in either the $\ket{\uparrow}$ or  $\ket{\downarrow}$ electron spin manifolds there is no spin superposition of the electron wavefunction and that the electron spin remains separable from the nuclear spins.


\textbf{Universal Control of $\text{2}^\text{N+1}$-dimensional Hilbert Space.}
Having identified a system for storing quantum information we show the condition under which such a system allows for arbitrary quantum operations.  A necessary and sufficient condition for complete controllablity, or universality, requires the nested commutators of the natural Hamiltonian Eq. \eqref{HSimple} and each of the control Hamiltonians are a closed Lie group $\text{SU}(2^{N+1})$ \footnote{In classical control theory of linear systems this is known as the Kalman rank condition}\cite{RabitzControl,SchirmerPRA}.  An equivalent diagrammatic representation relies on graph connectivity for assessing the controllability of quantum systems represented as Lie algebras \cite{AltafiniControlGraph,RabitzControlGraph}.

\begin{figure}[htbp]
\begin{center}
	\subfigure[1e-1n isotropic]{\includegraphics[scale=.50]{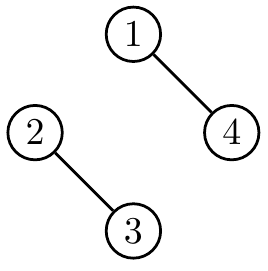}}
	\subfigure[1e-1n anisotropic]{\includegraphics[scale=.50]{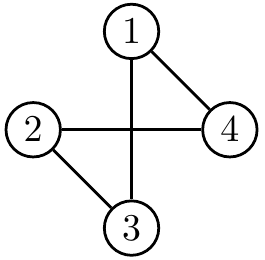}} 
	\subfigure[1e-3n fully anisotropic]{\includegraphics[scale=.50]{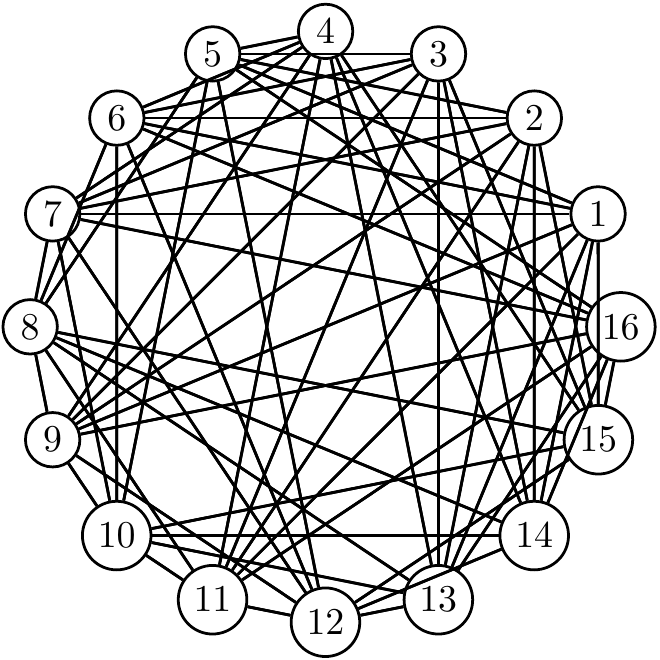}}

\caption{The connectivity of energy levels of 1e-Nn systems represented as $2^{N+1}$ node graphs.  
An edge is drawn between two nodes if the control Hamiltonian operator has a non-zero matrix element for the eigenstates represented by the nodes ($\bra{k}\hat{S}_x\ket{j} \neq 0$).  When the hyperfine interaction between \textit{any one} nuclear spin and the electron is purely isotropic universality is not achieved as in (a).}
\label{Connectivity}
\end{center}
\end{figure}

Our controls are time-dependent microwave fields oscillating at the electron spin resonance frequency ($\sim\beta_eg_{zz} B_0$) and parametrized by three values: $B_1^e$, the amplitude of an oscillating magnetic field ($\perp B_0 \hat{z}$), $\Omega$ the frequency of oscillation, and $\phi$ the phase of the oscillation:
\begin{equation}
\label{eq:controlHam}
\begin{array}{ll}
\ham_{\text{C}} (B_1^e, \Omega, \phi)  = & \\
\beta_e \text{g}_{zz} B_{1}^e(t) \Big( \cos \big(\Omega t + \phi(t)  \big) ~ \hat{S}_x ~ +   \sin \big (\Omega t + \phi(t)  \big) ~ \hat{S}_y \Big) &
\end{array}
\end{equation}
Barring degenerate eigenvalues $E_j$ of $\ham_0$ or degenerate transition frequencies ($\hbar \omega_{jk} = E_j - E_k$), the matrix representation of $\ham_0$ is \textit{strongly regular}.  As shown in \cite{AltafiniControlGraph}, this and the complete connectivity of the graph generated by the matrix elements of $\ham_\text{C}$ guarantee universality.  For the 1e-N nuclear spin system, distinct gyromagnetic ratios and hyperfine couplings for each nuclear spin guarantee the non-degeneracy of the eigenstates.  The hyperfine couplings and the Zeeman frequencies must also be chosen such that $\omega_{jk}/\omega_{j^\prime k^\prime} \neq 1$.  Lastly, the anisotropy of \textit{each} hyperfine interaction assures the complete connectivity of the graph given the form of $\ham_0$ and $\ham_\text{C}$. Figure \ref{Connectivity} shows several diagrammatic examples.  For universal control over a set of nuclear spins with resolved anisotropic hyperfine coupling to one electron spin, it is sufficient to apply only an amplitude modulated waveform to any electron spin transition at a fixed $\Omega$.

 We can provide some insight into how modulation of the electron spin state through shaped microwave fields provides control over the nuclear spins.  Flipping the electron spin changes the quantization axis of the nuclear spin.  Since the these two quantization axes are separated by a large angle, a sequence of spin evolutions under these two non-commuting axes permits arbitrary nuclear rorations.  If we consider collective motions of the N nuclear spins relative to the two electron spin states, this generates the complete algebra in $\text{SU}(2^{N+1})$.
Given the complexity of the full dynamics of these $2^N$ vectors, we can use optimal control methods developed for and applied to liquid state nuclear magnetic resonance (NMR) to engineer arbitrary unitaries \cite{KhanejaGRAPE,SMP_1,SMP_2} .
We limit the control fields of the electron-nuclear system to only the electron spin flip transitions and achieve quantum gates whose operation times are faster than if we relied upon nuclear spin nutation.  The idea of using pulsed ESR to quickly generate nuclear spin coherence and echoes was first used for spectroscopy \cite{SchweigerNuclearSpinEchoes,SchweigerNCTEchoes};  here, we extend this idea to not only generate nuclear coherence, but also suppress all other closed-system dynamics to generate a desired quantum gate.  For organic crystal systems of a few nuclear spins the values of the hyperfine interaction are tens of MHz.  For pulsed ESR systems the Rabi frequencies for an electron spin are also tens of MHz \cite{FreedMalonic,SchweigerDarkResonance,SchweigerBook,BrukerSpinReport}, while the nuclear Rabi frequency is at most 1 MHz \cite{KentgensMicroCoil}.  The nuclear-nuclear dipolar coupling is tens of kHz.


\textbf{The 1e-1n system.}
We demonstrate the utility of this control scheme by exploring Ramsey fringes \cite{Ramsey} and Hahn echoes \cite{HahnEcho} in a 1e-1n system \nobreakdash --- a single crystal of x-ray irradiated malonic acid	 \cite{McConnellMalonic,McCalley_Malonic,KangMalonic}.
The parametrized Hamiltonian is
$\ham_0/2 \pi =  \nu_s \hat{S}_z - \nu_n \hat{I}_z + A_{zx} \hat{S}_z \hat{I}_x + A_{zz} \hat{S}_z \hat{I}_z $ where
$\nu_s = 11.885$ GHz, $\nu_n = 18.1$ MHz, $A_{zx} \approx 14.2$ MHz, and $A_{zz} \approx -42.7$ MHz.  The control Hamiltonian Eq. \eqref{eq:controlHam}
 parameters are $\textsf{max}\big(\beta_e \text{g}_{zz} B_1^e(t)\big)/2\pi = 7$ MHz and $\Omega/2\pi = 11.909$ GHz.  No arbitrary phase controls $\phi(t)$ were used.  All experiments were performed on a home-built pulsed ESR spectrometer with a 12W solid-state microwave amplifier capable of amplitude modulations of 250 Ms/s using an  arbitrary waveform generator (AWG) and double-balanced mixer. A home-built low-temperature probehead cools the sample to 77K and contains a loop gap resonator with a volume of $\sim 55 \text{mm}^3$ and a Q of 250.

\textbf{Experimental Results.}
To implement an arbitrary unitary propagator we use the GRadient Ascent Pulse Engineering (GRAPE) \cite{KhanejaGRAPE} algorithm for finding the control field $\mathscr{H}_{\text{C}}$.  Constraints on the the modulation sequence, such as maximum nutation rate and pulse bandwidth, were chosen in accordance with our hardware limitations such as finite power amplifiers, modest AWGs, and finite bandwidth components. The simulated gate fidelity \cite{SMP_1} of useful gates is at least 0.98.

The equilibrium state of the ensemble system, $\rho_{\text{thermal}} = -\hat{S}_z$, has no net nuclear spin polarization, so we first transfer the available electron spin polarization to the nuclear spins.  This is achieved by selectively inverting the levels $\ket{2}$ and $\ket{4}$ or  $\ket{1}$ and $\ket{3}$ (see Figure \ref{fig:energy}).  We created coherence between nuclear eigenstates with an engineered nuclear $\pi/2$ pulse selective for only one of the electron manifolds, $U_{12}(\pi/2)$ \footnote{$U_{jk}(\theta) = e^{(-i\frac{\theta}{2} \sigma_x^{jk})}$. The operator $\sigma_x^{jk} = \ket{j}\bra{k} + \ket{k}\bra{j}$}.  The Ramsey fringe experiment measures the phase evolution under $\ham_0$.
 We halt evolution by again applying $U_{12}(\pi/2)$ and then transferring the polarization back to the electron spin.  By monitoring the relative amplitude of the electron spin echo at different times, $\tau$, we indirectly observe the nuclear spin dynamics.   If we introduce a refocusing pulse for both nuclear spin states, $U_r = U_{12}(\pi) \oplus U_{34}(\pi)$, after a time $\tau/2$, the coherent phase oscillations are refocused.

 
  \begin{figure}[htbp]    \centering
   \includegraphics[scale=.5]{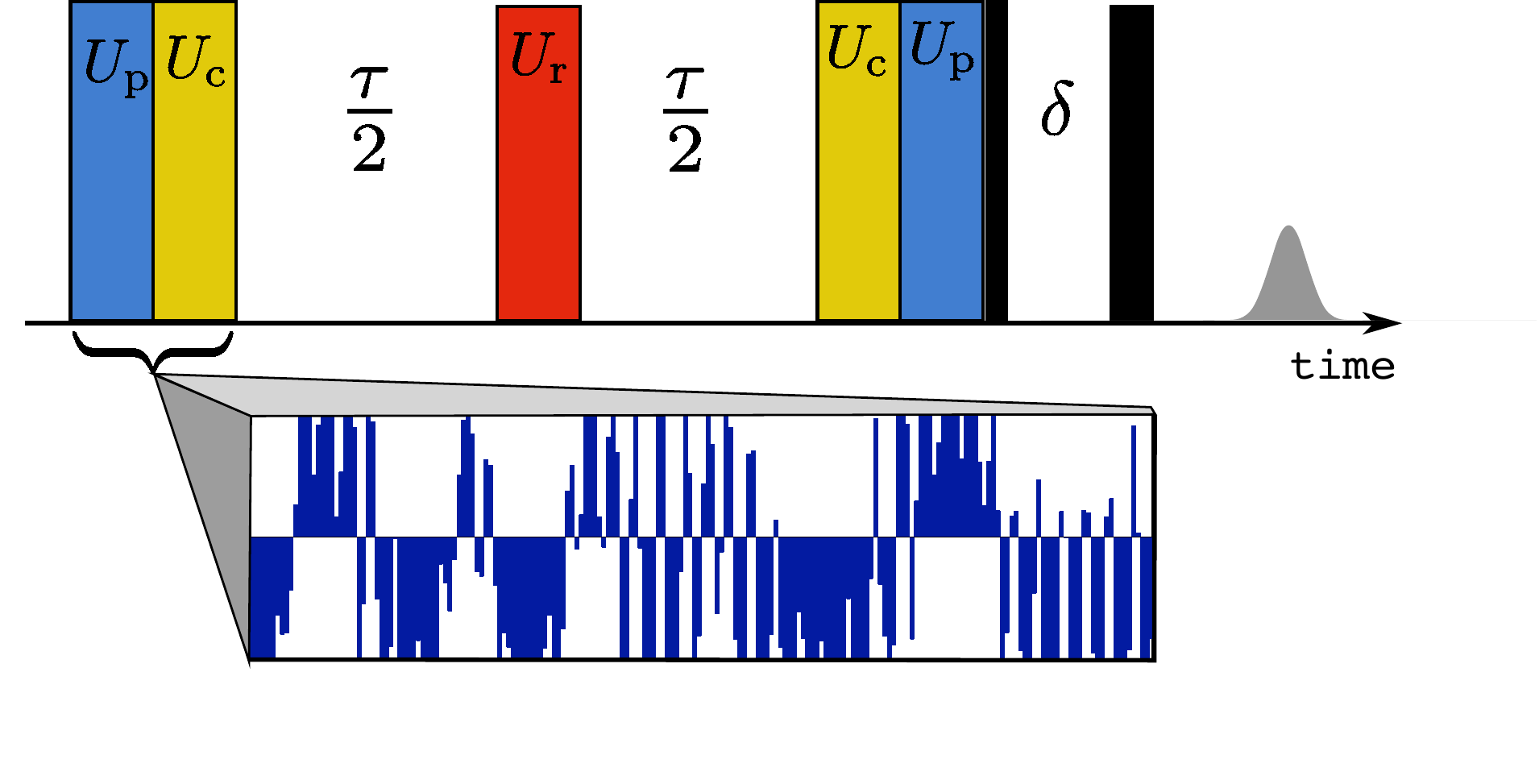} 
   \caption{Color online. Schematic pulse sequence for measuring Ramsey fringes and Hahn echoes.  $U_\text{p}$ creates a non-equilibirum population difference between levels 1\&2 and $U_\text{c}$ creates a coherence between nuclear spin in the S=-1/2 manifold.  During $\tau$, this coherence evolves under $\mathscr{H}_0$, acquiring an observable phase.  The coherence is transformed back to a population difference between nuclear spin levels and then to electron spin levels.  A pair of short, unmodulated pulses are used to detect an electron spin echo whose height is proportional to the resultant electron spin population.  With a refocusing pulse ($U_\text{r}$) the acquired phase is unraveled, leaving no modulation of the echo signal.  The waveform used to implement $U_\text{p}U_\text{c}$ is shown inset.  Note that all pulses are applied resonant with the 1-4 transition and induces transitions between 1-4, 2-4, 1-3, and 2-3 due to selection rules.}
   \label{fig:pulsesequence}
\end{figure}
  
\begin{figure}[htbp] 
    \centering
   \includegraphics[width=3.5in]{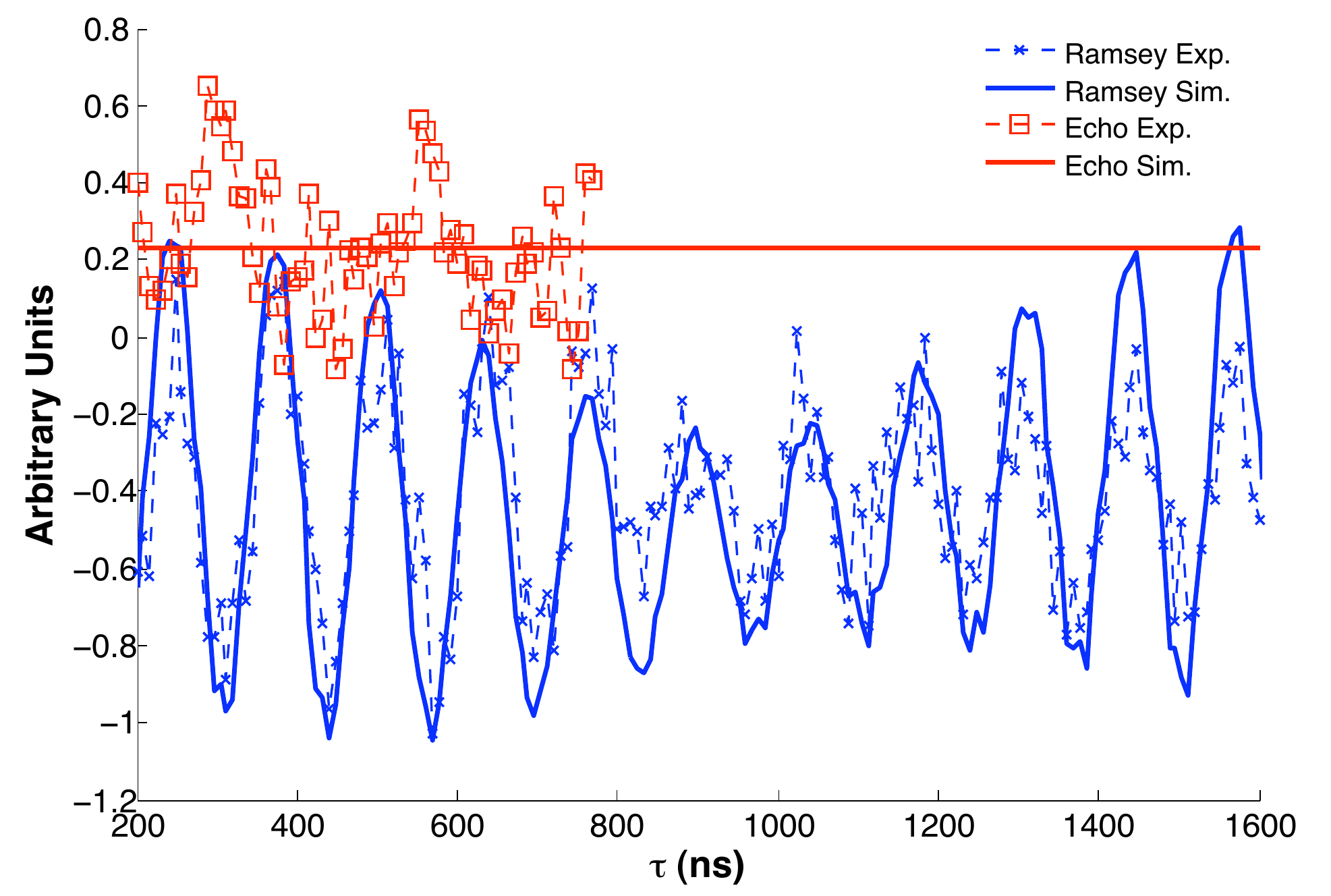} 
   \caption{Color online. Measurements of the electron spin echo as a function of $\tau$ between coherence transfer indirectly reveal the nuclear precession rate.  Numerical simulations of the experiment show agreement with the observed signal (signal-to-noise ratio $\approx$ 4.9).  The Ramsey fringe experiment ($\times$) reveals a clear precession of the coherence between the $\ket{1}$ and $\ket{2}$ states at roughly $8$ MHz .  For the echo data ($\square$), small pulse errors and incomplete phase cycling result in a small systematic oscillation at half the precession frequency; ideally the echo should be independent of  $\tau$.}
   \label{fig:ramsey12}
\end{figure}

 Figure \ref{fig:ramsey12} shows the coherent oscillations between nuclear coherence on levels $\ket{1}$ \& $\ket{2}$.   We implement the net unitaries, $U_{\text{pc}} =  U_{12}(\pot)U_{24}(\pi)$ and  $U_{\text{pc}}^{-1}$ as a single modulation sequence with total time (T) of 800ns and simulated fidelities (F)  0.99 and 0.98 respectively.  The nuclear $\pi$ pulse, $U_r$, was implemented in a time period (F=0.98, T=520ns) much shorter than would have been possible by addressing the nuclear spin transitions directly.  Again, it is key to realize that the nuclear pulse is achieved through modulation of the hyperfine interaction and is applied at the electron spin resonance frequency. 
 Simulations of the modulation sequences using our model 1e-1n Hamiltonian show agreement of the observed oscillation.


 \textbf{Conclusions and Outlook.}
 Using shaped pulse techniques for engineering quantum gates, we have described a method to control nuclear spins coupled via the hyperfine interaction to a nearby localized electron spin impurity using \textit{only} electron spin transitions.  Furthermore we have experimentally demonstrated the viability of these ideas using a testbed system of one electron spin (S=1/2) and one nuclear spin (I=1/2).  
 
 Using the same experimental setup this method can, in principle, extend to any number of nuclear spins that have resolvable anisotropic hyperfine interactions.  Specifically, we have found a 1e-2n \textsc{CNOT} gate, (see supporting materials for details) between the carbon and proton nuclear spins for malonic acid isotopically labeled with a $^{13}$C at the methylene position\cite{McConnell_13CMalonic}.  This gate can be performed in only 2$\mu$s.  If only nuclear nutation frequencies and nuclear dipole-dipole interactions were used, such a \textsc{CNOT} would take much longer due to the relative strength of $\mathbf{D}$ to $\mathbf{A}$.  Modulations of the anisotropic hyperfine interactions can generate quantum gates between nuclear spins without changing the state of or entangling the electron spin.  Previously studied solid-state spin systems with multiple resolved nuclear spins \cite{MehringSpinBus,JelezkoSingleESR,JelezkoCROT,ChildressNV,DuttNV} can thus be controlled by only modulating the electron spin states.
Moreover, it should be possible to augment the controllable Hilbert space by coupling localized electron spin states.  Such coupling may be electrical, with systems like phosphorous-doped silicon \cite{SchweigerSilicon} or optical in systems akin to nitrogen-vacany defects in diamond \cite{ChildressRepeater}.  As the ultimate utility of these systems depends on their decoherence, characterization of these mechanisms is key.  Precise unitary engineering can also be used to measure relaxation processes or perform quantum process tomography  in hyperfine coupled systems in a more controlled manner than has been reported \cite{FreedMalonic}. 
 
\section{Acknowledgments}
We thank Colm Ryan for providing his implementation of the GRAPE algorithm and Peter Allen for technical assistance in the experimental setup. We also thank Navin Khaneja for useful discussions; he recently provided a similar construction for control \cite{NavinHF}.  This work was supported in part by the National Security Agency (NSA) under Army Research Office (ARO) contracts DAAD190310125 and W911NF-05-1-0459 and by DARPA. JCY acknowledges support of a QUACGR fellowship from the Army Research Office (ARO).
\bibliography{ESR_RefsNoURL}

\onecolumngrid
\newpage
\section{Supporting Materials}
\begin{figure}[htbp] 
\begin{center}
   \includegraphics[scale=.5]{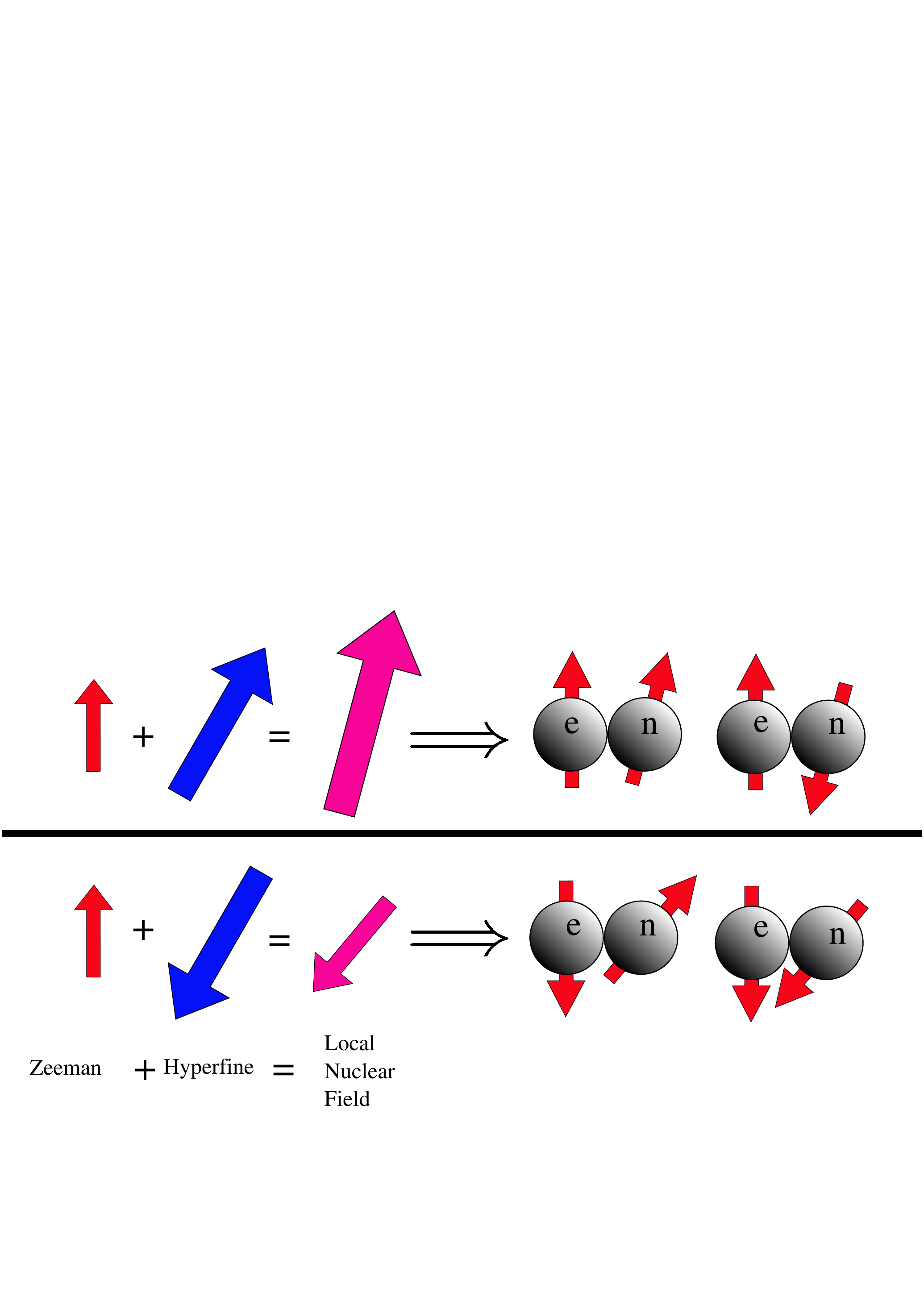} 
   \caption{Contributions to nuclear field. The nuclear Zeeman field adds to the local hyperfine field yielding the quantization axis for the nuclear spin.  Because the electron spin can be aligned or anti-aligned with the main magnetic field, the local hyperfine field can be either positive or negative.  The nuclear spin aligns (or anti-aligns) with the local magnetic field.  The presence of an anisotropic term hyperfine interaction comparable to the nuclear Zeeman interaction prevents the nuclear spins aligning parallel or antiparallel to the electron spins.  Note that the quantization axis of the nuclear spin differs by a finite angle in these two cases.}
   \label{fig:localhf}
   \end{center}
\end{figure}

\begin{figure}[htbp] 
\begin{center}
	\subfigure[Time Domain]{\includegraphics[scale=.40]{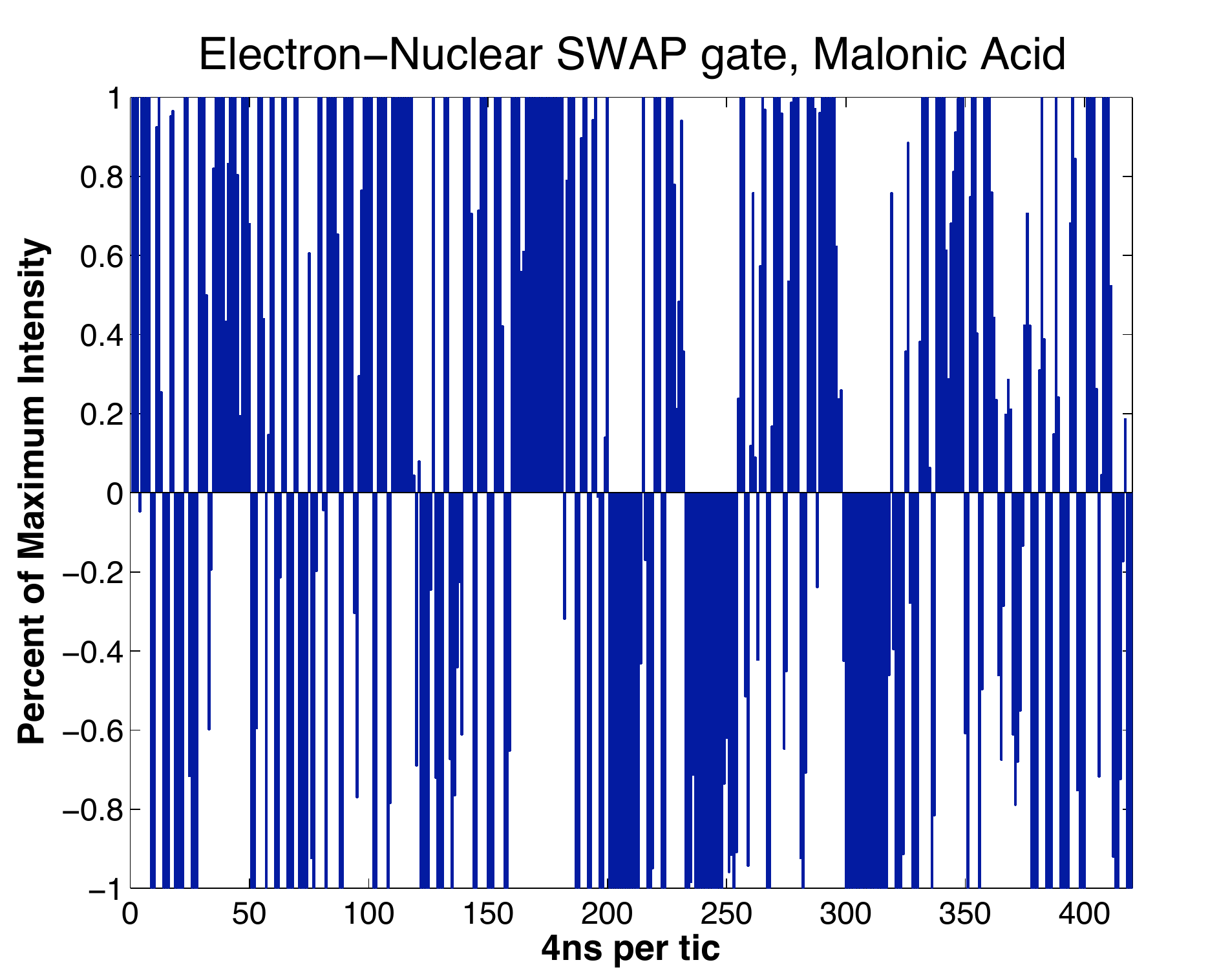}}   
	\subfigure[Frequency Domain]{\includegraphics[scale=.40]{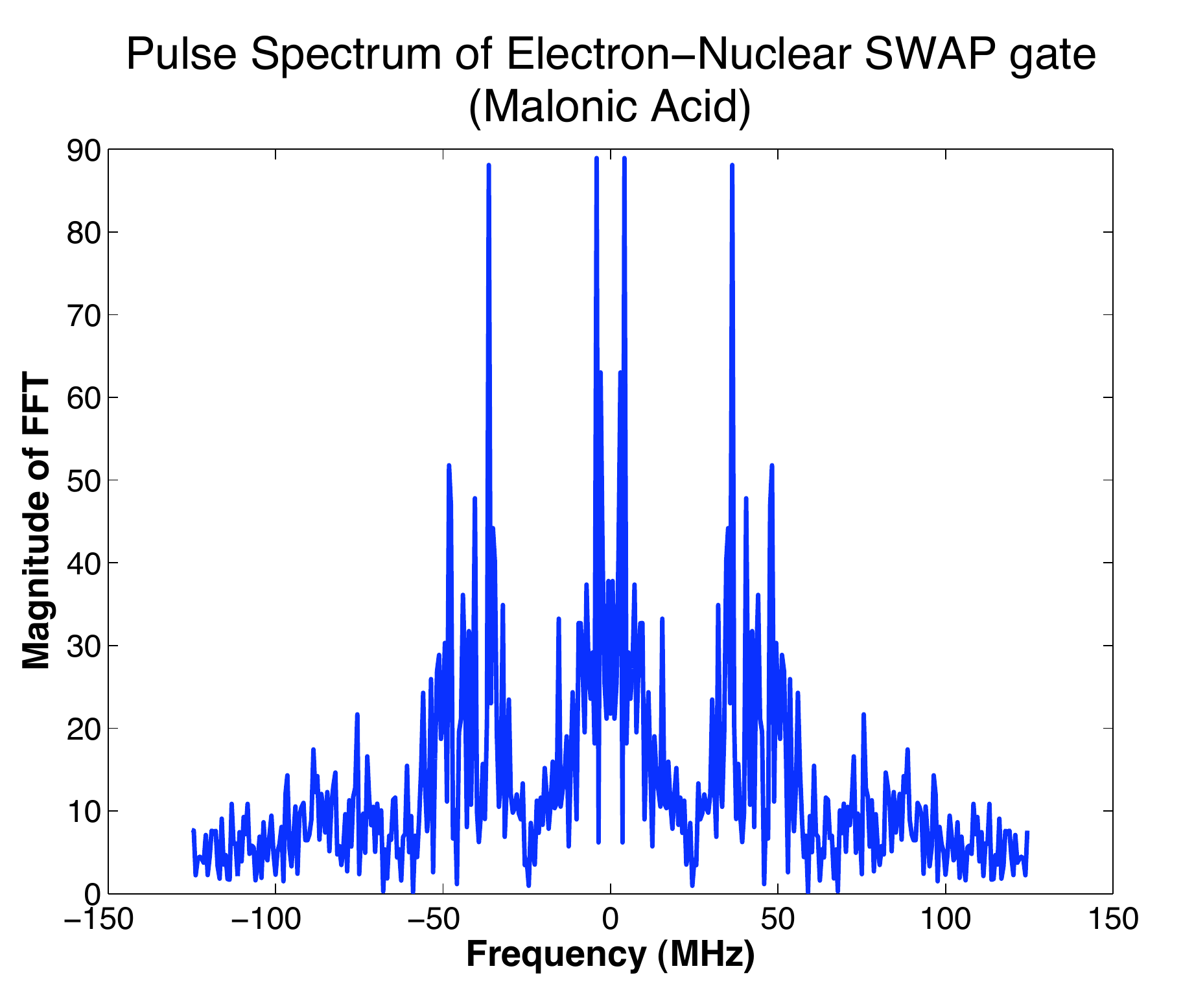}}   

	\caption{Electron-Nuclear SWAP gate.  For the same field and crystal orientation used in the above experiment, we find a modulation sequence (a) to perform a \textsc{SWAP} gate between the electron and the nuclear spin.  The sequence consists 420 4ns intervals each specifying a pulse amplitude from -1 to 1, corresponding to $\pm \omega_1^e$ the electron spin Rabi frequency.  This gate has a simulated fidelity of $0.999$ and has been experimentally implemented.  The frequency domain (b) shows the spectral components of the sequence.  Note that nearly half of the power is contained outside the $\pm 25$MHz bandwidth centered about 0.  If we take into account the Q of the resonator, thus attenuating these higher frequency components, the simulated fidelity degrades to $0.964$.}
   \label{fig:en_swap}
   \end{center}
\end{figure}

\begin{figure}[htbp] 
\begin{center}
	\includegraphics[scale=0.85]{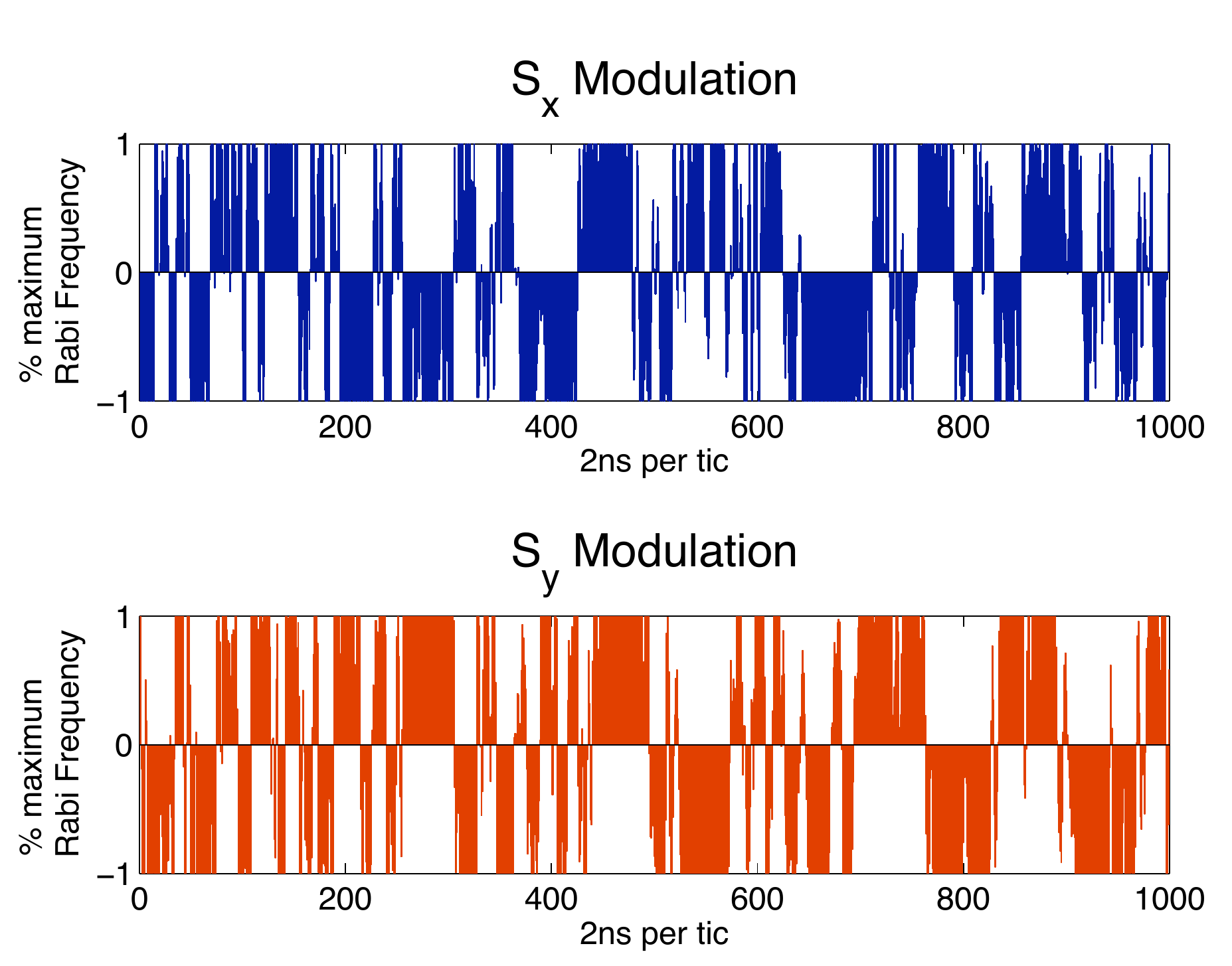}
	\caption{Nuclear-Nuclear \textsc{CNOT} gate.  If the carbon nucleus closest to the paramagnetic defect of malonic acid is a $^{13}$C isotope (I=1/2), we have a 1e-2n system.  Using hyperfine tensor values reported in the literature, we find a modulation sequence to implement a \textsc{CNOT} gate between two nuclear spins: $U_\text{1e-2n} = \mathbf{1} \otimes \text{C}_H \text{NOT}_C$.  At the same crystal orientation used above, we find a sequence modulating both the amplitude and the phase of the control Hamiltonian with 1000 2ns intervals and a maximum Rabi frequency of 15MHz. The time for performing the gate using nuclear-nuclear dipole couplings (tens of kHz) in the absence of an electron spin would be much longer. In our scheme the nuclear-nuclear dipole couplings enter as a source of decoherence and can be included in the model to generate gates robust to this interaction.}
   \label{fig:en_swap}
   \end{center}
\end{figure}

\begin{figure}[htbp] 
\begin{center}
   \includegraphics[scale=.5]{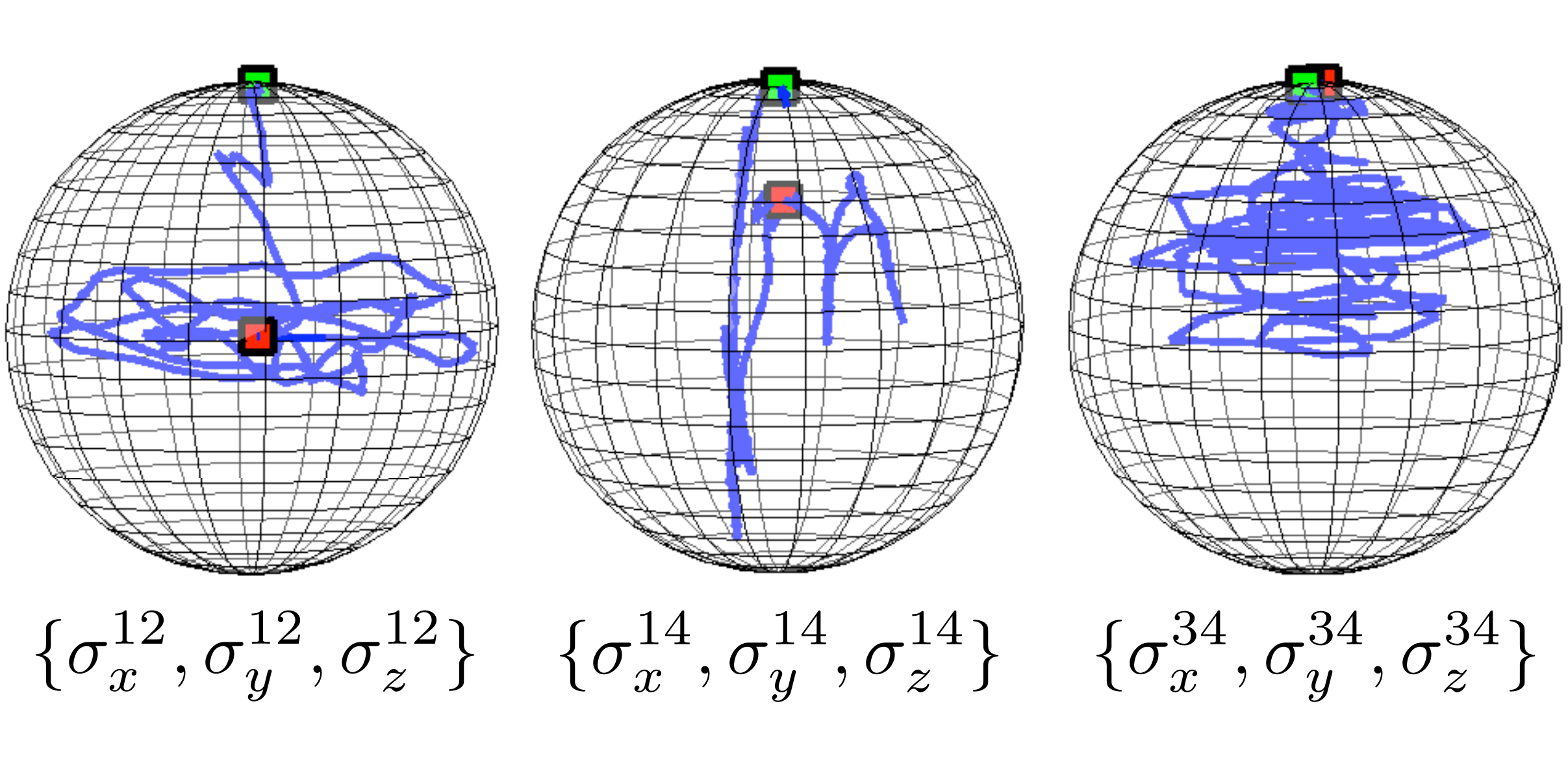} 
   \caption{Trajectories of spin systems on projected Bloch Spheres.  After a polarization transfer from the equilibrium electron spin state to the nuclear spin, we apply a $90^\circ$ rotation between the $\ket{1}$ and $\ket{2}$ manifold.  The trajectory of the spin vector can be traced by taking an equivalent Bloch sphere between any two levels, where the usual operators $\{\sigma_x,\sigma_y,\sigma_z\}$ are analogous to  $\sigma_x^{jk} = \ket{j}\bra{k} + \ket{k}\bra{j}$,  $\sigma_y^{jk} = i\ket{j}\bra{k} - i \ket{k}\bra{j}$, and $ \sigma_z^{jk} = \ket{j}\bra{j} - \ket{k}\bra{k}$. For the 12-pair, the state begins along the $\hat{z}$ axis (green) and then rotates down to the equatorial plane (red).  The 14-pair reduces it's $\hat{z}$ component by a factor of two, as half of the state goes into $\sigma_y^{12}$ coherence; the 34-pair begins and ends along the $\hat{z}$-axis, as the rotation is in a subspace orthogonal to 34.  Although the trajectories the spin takes can be very complicated, only the starting and end points are significant.}   \label{fig:blochsphere}
   \end{center}
\end{figure}

\end{document}